\documentclass[aps,amsfonts,amsmath,prd,preprint,nofootinbib,tightenlines]{revtex4}
\usepackage{epsf}

\def\be{\begin{equation}}
\def\ee{\end{equation}}
\def\bea{\begin{eqnarray}}
\def\eea{\end{eqnarray}}

\def\obs{{\text{obs}}}

\def\Jchoosen{\left(\begin{array}{c}
J \\ n \\
\end{array} \right)}

\def\erfc{{\mathop{\text{erfc}}}}

\begin{document}

\title{Anthropic prediction in a large toy landscape}
\author{Ken D. Olum and Delia Schwartz-Perlov}
\affiliation{Institute of Cosmology, Department of Physics and Astronomy\\
Tufts University, Medford, MA 02155, USA }

\begin{abstract}
The successful anthropic prediction of the cosmological constant
depends crucially on the assumption of a flat prior distribution.
However, previous calculations in simplified landscape models showed
that the prior distribution is staggered, suggesting a conflict with
anthropic predictions. Here we analytically calculate the full
distribution, including the prior and anthropic selection effects,
in a toy landscape model with a realistic number of vacua, $N \sim
10^{500}$. We show that it is possible for the fractal prior
distribution we find to behave as an effectively flat distribution
in a wide class of landscapes, depending on the regime of parameter
space.  Whether or not this possibility is realized depends on
presently unknown details of the landscape.
\end{abstract}

\maketitle

\section{Introduction}

The observed value of the cosmological constant $\Lambda$ is about
$120$ orders of magnitude smaller than theoretically
expected\footnote{Here and below we use reduced Planck units,
$M_{RP} \equiv M_p^2/8\pi =1$, where $M_p$ is the Planck mass. The
theoretical expectation could be as ``large'' as $10^{-56}$, for
example because of supersymmetry.  However, a discrepancy of
$56$ orders of magnitude still needs to be explained.} \be
\Lambda_0\sim 10^{-120}. \ee

One explanation for this observation assumes that $\Lambda$ is an
environmental parameter which has different values in different
parts of the ``multiverse"
\cite{Weinberg87,Linde87,AV95,Efstathiou,MSW,GLV,Bludman,AV05}. The
probability for a randomly picked observer to measure a given value
of $\Lambda$ can then be expressed as \cite{AV95}
\be
P_\obs(\Lambda)\propto P(\Lambda)n_\obs(\Lambda), \label{Pobs}
\ee
where $P(\Lambda)$ is the prior distribution or volume fraction of
regions with a given value of $\Lambda$ and $n_\obs(\Lambda)$ is the
anthropic factor, which is proportional to the number of observers
that will evolve per unit volume. If we assume that $\Lambda$ is the
only variable ``constant'', then the density of observers is roughly
proportional to the fraction of matter clustered in large galaxies,
$n_\obs(\Lambda) \propto f_G(\Lambda)$. Using the Press-Schechter
approximation \cite{Press:1973iz} for $f_G(\Lambda)$ we can write
\cite{astro-ph/0611573}
\be
\label{eqn:erfc} n_\obs(\Lambda) \sim
\erfc\left[\left(\frac{\Lambda}{\Lambda_c}\right)^{1/3}\right]
\ee
where we have normalized $n_\obs$ to be 1 for $\Lambda = 0$. We have
parameterized the anthropic suppression with a value $\Lambda_c$,
which depends on such things as the amplitude of primordial
fluctuations and the minimum size of galaxy that can contain
observers.  For the parameters used in Refs.
\cite{astro-ph/0611573,astro-ph/0410281}, $\Lambda_c$ is about 10
times the observed value of $\Lambda$,
\be
\Lambda_c \sim 6\times
10^{-120}.
\ee

For the qualitative arguments and toy model of the present paper,
the precise value of $\Lambda_c$ will not matter, and we can use a
Gaussian instead of $\erfc$, to get
\be
\label{eqn:nobs}
n_\obs(\Lambda) = e^{-(\Lambda/\Lambda_c)^{2/3}}
\ee

The prior distribution $P(\Lambda)$ depends on the unknown details
of the fundamental theory and on the dynamics of eternal inflation.
However, it has been argued \cite{AV96,Weinberg96} that it should be
well approximated by a flat distribution,
\be
P(\Lambda)\approx {\rm
const} \label{flat},
\ee
because the window where $n_\obs(\Lambda)$
is substantially different from zero, is vastly less than the
expected Planck scale range of variation of $\Lambda$.  Any smooth
function varying on some large characteristic scale will be nearly
constant within a relatively tiny interval. Thus from Eq.\
(\ref{Pobs}),
\be
P_\obs(\Lambda)\propto n_\obs(\Lambda).
\label{PnG}
\ee

Indeed the observed value of $\Lambda$ is reasonably typical of
values drawn from the distribution of Eq.\ (\ref{eqn:erfc}). This
successful prediction for $\Lambda$ depends on the assumption of a
flat volume distribution (\ref{flat}).  If, for example, one uses
$P(\Lambda) \propto\Lambda$ instead of (\ref{flat}), the $2\sigma$
prediction would be $\Lambda/\Lambda_0 <500$, giving no satisfactory
explanation for why $\Lambda$ is so small \cite{Pogosian}.

A specific class of multiverse models is given by the landscape of
string theory \cite{BP,Susskind,AHDK}.  In such models there are of
order $10^{500}$ different vacua with various cosmological constants
\cite{Douglas,AshokDouglas,DenefDouglas}. The dynamics of eternal
inflation populates the multiverse with all possible vacua (or
bubbles) by allowing for the nucleation of one vacuum within the
other, according to transition rates which determine the probability
of going from one vacuum to another.

Given a specific string theory landscape, we would like to be able
to predict the cosmological constant that we should expect to
observe according to Eq.~(\ref{Pobs}). We will assume the prior
probability $P(\Lambda)$ is given by the relative bubble abundances
of different vacua. Since an eternally inflating multiverse contains
an infinite number of each type of vacuum allowed in the landscape,
it is necessary to use some regularization procedure to compute the
prior probability distribution.  Many such regularization
procedures, or probability measures, have been proposed
\cite{LLM94,Bousso,Alex,Aguirre,Linde07A}. For a more complete and
up to date account see Ref. \cite{Linde07B} and the references
therein. Here we will use the pocket-based measure introduced in
Refs.\ \cite{GSPVW,ELM}. Refs.\ \cite{SPV,SP} computed prior
probabilities\footnote{Strictly speaking bubble abundances were
calculated.} of different vacua in toy models \cite{BP,AHDK} and did
not find a smooth distribution of possible cosmological constants.
Instead, for the specific models and parameters they studied, there
were variations of many orders of magnitude in the prior
probabilities of different vacua.  However, to allow for numerical
solution, Refs.\ \cite{SPV,SP} used models with a relatively small
number of vacua and worked only in a first-order approximation.

Here we will consider a toy model in which transition probabilities
are computed as though all changes in $\Lambda$ were by some fixed
amount $c$.  In this simple model, we can analytically study
probability distributions for a realistic number of vacua, $N \sim
10^{500}$. We find that when $c$ is around 1, there is a smooth
distribution of vacua in the anthropic range, and the anthropic
prediction of Eq.~(\ref{PnG}) applies.  But when $c$ is smaller by a
few orders of magnitude, the behavior is very different. In this
case, the $P(\Lambda)$ factor is more important than
$n_\obs(\Lambda)$ in Eq.~(\ref{Pobs}). Thus we would expect to live
in a region with large $\Lambda$, and so only a few galaxies. In
such a multiverse, the anthropic procedure would not explain the
observed small value of $\Lambda$.

The plan of this paper is as follows: In section \ref{abundances} we
will outline the method used to calculate bubble abundances.
Additional details of the bubble abundance calculation are presented
in the appendix.  We will then define our model in section
\ref{toymodel}, and calculate the prior probability distribution. In
section \ref{fulldistribution} we will investigate the behavior of
$P_{obs}(\Lambda)$. We end with a discussion in section
\ref{discussion}.

\section{Bubble abundances}
\label{abundances}

We review here the procedure for calculating the volume fraction of
vacua of a given kind, using the ``pocket-based measure'' formalism of
Refs.\ \cite{GSPVW,SPV}.

Vacua with $\Lambda\leq 0$ are said to be terminal.  There are no
transitions out of them.  Vacua with $\Lambda>0$ are recyclable.  If
$j$ labels such a vacuum, it may be possible to nucleate bubbles of a
new vacuum, say $i$, inside vacuum $j$.
The transition rate $\kappa_{ij}$ for this process is defined as the
probability per unit time for an observer who is currently in vacuum
$j$ to find herself in vacuum $i$. Using the logarithm of the scale
factor as our time variable,
\be
\kappa_{ij}=\Gamma_{ij}\frac{4\pi}{3}H_j^{-4}, \label{kappa}
\ee
where $\Gamma_{ij}$ is the bubble nucleation rate per unit physical
spacetime volume (same as $\lambda_{ij}$ in \cite{GSPVW}) and
\be
H_j = (\Lambda_j/3)^{1/2} \label{Hj}
\ee
is the expansion rate in
vacuum $j$.

Transition rates depend on the details of the landscape. However, if
$\Lambda_i < \Lambda_j$, the rate of the transition upward from $i$
to $j$ is suppressed relative to the inverse, downward transition,
by a factor which does not depend on the details of the process
\cite{EWeinberg}\footnote{We assume only Lee-Weinberg tunnelings and
do not consider Farhi-Guth-Guven (FGG) tunnelings \cite{FGG}. These
FGG tunnelings may be faster in upward transition rates, but their
interpretation is unclear \cite{AJ,AGJ}, and the resulting spacetime
cannot be directly handled by the ``pocket-based''
measure we employ here.}, \be
\label{eqn:updown} \kappa_{ji} = \kappa_{ij} \exp \left[-24
\pi^2\left(\frac{1}{\Lambda_i}-\frac{1}{\Lambda_j}\right)\right] \ee

Given the entire set of rates $\kappa_{ij}$, we can in principle
compute the bubble abundance $p_{\alpha}$ for each vacuum $\alpha$,
following the methods of Refs.\ \cite{GSPVW,SPV}.  An exact
calculation would require diagonalizing an $N\times N$ matrix.  But
as in Ref.\ \cite{SPV}, we can make the approximation that all
upward transition rates are tiny compared to all downward transition
rates from a given vacuum (see also the appendix). In that
approximation, we can compute probabilities as follows.

First, define the total down-tunneling rate for a vacuum $j$,
\be
D_j = \sum_{\Lambda_i < \Lambda_j}\kappa_{ij}\,.
\ee

Then define the dominant vacuum, referred to as vacuum $*$, as that
recyclable vacuum whose $D_j$ is the smallest.  Since bubble
nucleation rates are suppressed in low-energy vacua, we expect
$\Lambda_*$ to be fairly small, however we would not expect it to be
so small as to be in the anthropic range. In Bousso-Polchinski and
Arkani-Hamed-Dimopolous-Kachru type landscapes \cite{BP,AHDK} it can
be shown that this vacuum will have no downward transitions to vacua
with positive $\Lambda$. To see that this is true, imagine that in
some direction $\Lambda_*$ can jump downward to
$\Lambda_{\alpha}>0$. Now if we compare $D_{\alpha}$ to $D_*$ we see
that each term contributing to $D_{\alpha}$ is less than the
corresponding term (i.e., the transition rate in the same direction)
in $D_*$ because $\Lambda_{\alpha}<\Lambda_*$ and jump sizes in the
same direction are the same.  This implies $D_{\alpha}<D_*$ which
contradicts our definition of $D_*$ as the vacuum with the smallest
sum of downward transition rates.

Once we have identified the dominant vacuum, the probability for any
vacuum $\alpha$ is given by (see Appendix)
\be
\label{eqn:pl}
p_\alpha = \sum \frac{\kappa_{\alpha a}\kappa_{ab}\cdots \kappa_{z*}}
{(D_a-D_*)(D_b-D_*)\cdots(D_z-D_*)}
\ee
where the sum is taken over all chains of intermediate
vacua $a,b,\ldots,z$ that connect the vacuum $\alpha$ to the dominant
vacuum.

\section{Toy model}
\label{toymodel}

\subsection{Model}

We will consider a toy version of the Arkani-Hamed-Dimopolous-Kachru
(ADK) model \cite{AHDK}. We let there be $2J$ directions and so $N =
2^{2J}$ vacua.  We will choose $J \approx 800$, so that $N\sim
10^{500}$.  Each vacuum can be specified by a list of numbers
$\{\eta_1,\ldots,\eta_{2J}\}$, where $ \eta_i = \pm 1$, and the
cosmological constant is
\be
\label{eqn:Lambda}\Lambda = \bar\Lambda
+ \frac{1}2\sum_i \eta_i c
\ee
The ``toy'' feature of this model is that all jumps have the same
size, $c$.

We will take the average cosmological constant $\bar\Lambda$ to be in
the range $(0, c)$.  All vacua with J $+$ coordinates and $J$ $-$
coordinates will have $\Lambda=\bar\Lambda$.  Each vacuum has $2J$
neighbors to which it can tunnel by bubble nucleation.  Each
nucleation event either increases or decreases the cosmological
constant by $c$.

The model as given above is of no use for anthropic reasoning. Vacua
exist only with widely separated cosmological constants,
$\bar\Lambda,\bar\Lambda +c$\ldots, therefore we would not expect any
in the anthropic range.  So we will modify the model by artificially
perturbing the $\Lambda$ of each vacuum to produce a smooth number
distribution.  Vacua originally clustered at $\bar\Lambda$ will be
spread out over the range from 0 to $c$.  This will cover the
anthropic range of vacua with $\Lambda > 0$, and so if the vacua are
dense enough we will find some anthropic vacua.  We will not, however,
take account of these perturbations in computing probabilities.

We will only be interested in the vacua near $\Lambda = 0$, which
are those at $\bar\Lambda$ before the perturbation procedure above.
All these have exactly the same transition rates, and thus there is
no single dominant vacuum.  So, in addition to the ``smearing''
above, we will make a small perturbation to decrease the total
tunneling rate of some specific vacuum $\Lambda_*$, so that it is
the dominant one and the procedures of the last section can be
applied.

\subsection{Distribution of vacua}

The dominant vacuum, and all other vacua of interest have $J$ $+$
coordinates and $J$ $-$ coordinates.  Thus the other vacua are reached
from the dominant vacuum by taking an equal number of up jumps and
down jumps.  We will classify the vacua by a parameter $n$, the
minimum number of up jumps required to reach a given vacuum from the
dominant vacuum.  We will call $n$ the level of the vacuum.  Thus a
vacuum of level $n$ differs from the dominant vacuum in $2n$
coordinates, $n$ of which are $+$ where the dominant vacuum had $-$,
and another $n$ vice versa.

The total number of vacua of level $n$ is thus
\be
N_n =
\Jchoosen^2 =\left(\frac{J!}{n!(J-n)!}\right)^2 \label{eqn:Nn}
\ee
We imagine these to be smeared over a range $c$, so their density is
\be
\rho_n = N_n/c = 1/\Delta_n\,.
\ee

The likelihood that there is no vacuum in a range of size $x$ is
$\exp(-\rho_n x)$, thus the median $\Lambda$ of the lowest-$\Lambda$
vacuum is \be \label{eqn:Lambdan} \Lambda_n = (\ln2)/\rho_n =
c(\ln2)/N_n \ee We will take a typical realization to be one whose
lowest-$\Lambda$ vacuum is at this median position.  Above the
lowest-$\Lambda$ vacuum of level $n$, there are $N_n-1$ more with
higher $\Lambda$, with the typical interval in $\Lambda$ being
$\Delta_n$.  For a typical realization it is sufficient to take
these vacua as evenly spaced, so that they are at \be
\Lambda_{n,\ell} =\Lambda_n+(\ell-1)\Delta_n \label{eqn:lambdani}
\ee where $1 \leq \ell \leq N_n$.

\subsection{Probabilities}

We now use the formalism of Sec.\ \ref{abundances} to calculate the
relative abundances of different vacua in our toy model.

The relative abundance of each vacuum $\alpha$ is given by a sum
over all chains that connect it to the dominant vacuum, Eq.\
(\ref{eqn:pl}).  The minimum number of transitions in such a chain
is $2n$.  Longer chains can be formed by jumping one way and then
later the opposite way in the same direction.  But these chains will
have extra suppression factors because of the extra jumps, so the
probability will be accurately given by including only
minimum-length chains.  Furthermore, the paths that maximize the
bubble abundances are those that entail first making all the up
jumps and then following with a sequence of down jumps.  The reason
is that the up-jump suppression factor, Eq.\ (\ref{eqn:updown}), is
least when the starting $\Lambda$ for the jump is highest.  Thus it
is best not to jump down until one has made all the necessary
up-jumps. So we only consider the contribution of paths which
consist of making all upward jumps first and then following with
downward jumps.  In this case, we can reorganize Eq.\
(\ref{eqn:pl}),
\be
\label{eqn:pl2} p_\alpha = \sum
\frac{\kappa_{\alpha a}}{D_a-D_*}\frac{\kappa_{ab}}{D_b-D_*}\cdots
\frac{\kappa_{rs}}{D_{s}-D_*} \kappa_{st}
\frac{\kappa_{tu}}{D_t-D_*}\cdots \frac{\kappa_{z*}}{D_z-D_*}
\ee

The transition rates to the right of the factor $\kappa_{st}$ in Eq.
(\ref{eqn:pl}) are upward rates, and those to the left are downward
rates. $\kappa_{st}$ represents the first downward jump after having
made $n$ upward jumps from the dominant vacuum.

Now we will approximate $D_*\ll D_j$, since the transition rates are
suppressed for low $\Lambda$ vacua.  Furthermore, in our single jump
size model, all downward jumps from the same site have the same
transition rate, so for $\Lambda_i < \Lambda_j$,
\be
\label{eqn:Jj}
\frac{\kappa_{ij}}{D_j-D_*} \simeq \frac{1}{J_j}
\ee
where $J_j$ is the number of $+$ coordinates in vacuum $j$.

Using Eq.\ (\ref{eqn:updown}),
\be
\label{eqn:kappaDupdown}\frac{\kappa_{ji}}{D_j-D_*} =
\frac{\kappa_{ij}}{D_j-D_*}\exp \left[-24
\pi^2\left(\frac{1}{\Lambda_i}-\frac{1}{\Lambda_j}\right)\right]
\ee
The product of all such suppression factors is just
\be
\label{eqn:suppression} S = \exp \left[-24
\pi^2\left(\frac{1}{\Lambda_*}-\frac{1}{\Lambda_n}\right)\right]
\ee
where $\Lambda_n = nc$ is the maximum $\Lambda$ reached.

Factoring out the suppression factors, we find $n$ terms given by
Eq.\ (\ref{eqn:Jj}) for the up jumps.  The first one has $J_j=J+1$,
since there are $J$ $+$ coordinates in the dominant vacuum.  The
next has $J+2$, and so on up to $J+n$.  Thus the product is
\be
\frac{\kappa_{tu}}{D_t-D_*}\cdots \frac{\kappa_{z*}}{D_z-D_*} =
J!/(J+n)!
\ee
The down-jumps are similar, except that the one at $n$
is missing, giving
\be
\frac{\kappa_{\alpha a}}{D_a-D_*}\cdots
\frac{\kappa_{rs}}{D_s-D_*} = J!/(J+n-1)!
\ee
The up- and the
down-jumps can be taken in any order, so there are $(n!)^2$ equally
weighted paths to reach the same vacuum.  Thus the prior probability
of a vacuum of level $n$ is
\be
P_n \propto
(n!)^2 \left[\frac{J!} {(J+n)!}\right]^2(J+n)\kappa_{st} \exp \left[
 24\pi^2\left(\frac{1}{cn} - \frac{1}{\Lambda_*}\right)\right]
\label{eqn:pn1}
\ee
We set the first downward jump rate, which has
no canceling denominator,
\be
\label{eqn:kst} \kappa_{st} \approx
\exp \left[ - \frac{6 \sqrt{3c}\pi^2}{\Lambda_n^{3/2}}\right]= \exp
\left[ -\frac{6\sqrt{3}\pi^2}{c n^{3/2}}\right]
\ee
which is the
rate we would have for a Bousso-Polchinski(BP) type model \cite{SPV}
with $\Lambda_j \gg \Delta \Lambda$.

\section{Distribution for the observed $\Lambda$}
\label{fulldistribution}

Given the above, we are in a position to calculate the probability
of observing each value $\Lambda_{n, \ell}$ in a typical realization
of our toy model.  These are given by \be P_\obs(\Lambda_{n, \ell})
\propto P_n n_\obs(\Lambda_{n, \ell}) \ee The chance that we live in
a world of a given level $n$ is then given by \be P_\obs(n) \propto
P_n \sum_{\ell} n_\obs(\Lambda_{n, \ell}) \label{eqn:Pobsn} \ee

We will consider two cases.  When $\Lambda_n\ll\Lambda_c$, the sum
can be approximated by an integral, \be \label{eqn:littlesum}
\sum_{\ell} n_\obs(\Lambda_{n, \ell}) \approx \frac{1}{\Delta_n}
\int_0^\infty d\Lambda n_\obs(\Lambda) =
\frac{3\sqrt{\pi}\Lambda_c}{4\Delta_n} =
\frac{3\sqrt{\pi}N_n\Lambda_c}{4c} = \frac{3\sqrt{\pi}\Lambda_c\ln2}
{4\Lambda_n} \ee Including Eqs.\ (\ref{eqn:Nn}, \ref{eqn:pn1},
\ref{eqn:kst}), we find \be \label{eqn:Pobs1} P_\obs(n) \propto
\frac{3\sqrt{\pi}\Lambda_c}{4c}\left(\frac{J!^2}{(J-n)!(J+n)!}\right)^2
(J+n) \exp \left[\frac{24\pi^2}{cn}-\frac{6\sqrt{3}\pi^2}{c
n^{3/2}}\right] \ee

We have not included the term involving $\Lambda_*$, which is the
same for all $P_\obs(n)$.  In Eq.\ (\ref{eqn:Pobs1}), $P_\obs(n)$ is
a decreasing function of $n$.

On the other hand, when $\Lambda_n>\Lambda_c$, Eq.~(\ref{eqn:Pobsn})
will be dominated by the first term, and we can write \be
\label{eqn:bigsum} \sum_{\ell} n_\obs(\Lambda_{n, \ell}) \approx
e^{-(\Lambda_n/\Lambda_c)^{2/3}} \ee

Including Eqs.\ (\ref{eqn:pn1}, \ref{eqn:kst}), we find
\be
\label{eqn:Pobs2} P_\obs(n) \propto
\left(\frac{J!\,n!}{(J+n)!}\right)^2 (J+n) \exp
\left[\frac{24\pi^2}{cn}-\frac{6\sqrt{3}\pi^2}{c n^{3/2}}
-\left(\frac{\Lambda_n}{\Lambda_c}\right)^{2/3}\right]
\ee
In Eq.\
(\ref{eqn:Pobs2}), $P_\obs(n)$ increases with increasing $n$ while
$n$ is small and the last term in the exponent is dominant, but
it decreases when $n$ is larger and the other terms are dominant.

The division between regimes occurs when
$\Lambda_n\sim\Lambda_c$, i.e.,
\be
\Jchoosen^{-2}c\ln2 \sim
\Lambda_c\sim 6\times10^{-120}
\ee
With $c\sim 1$, we find $n\sim
34$.  The dependence on $c$ is weak, with $c\sim 10^{-3}$
corresponding to $n\sim 33$.  For $n$ in this range, changing $n$ by
one unit changes $\Lambda_n$ by a factor of about 500.  Thus there
is at most one $n$ with $\Lambda_n\sim\Lambda_c$.

If we compare Eq.\ (\ref{eqn:Pobs1}) and Eq.\ (\ref{eqn:Pobs2}) for
the same $n$, we see that they differ by a factor of $\Lambda_cN_n/c
\exp{(\Lambda_n/\Lambda_c)^{2/3}}\sim 1$ if $\Lambda_n \sim
\Lambda_c$, so there is no big jump due to switching regimes.

Now let us start with $n = 1$ and increase $n$.  Certainly with $n =
1$, $\Lambda_n\gg\Lambda_c$ by a huge factor, we are in the regime of
Eq.\ (\ref{eqn:Pobs2}), and $P_\obs$ is infinitesimal.  As we increase
$n$, $P_\obs$ increases.  Once $n$ is significantly above 1, we can
approximate the increase from one step to the next as
\be
\label{eqn:pp1}
\frac{P_\obs(n+1)}{P_\obs(n)}\approx \left(\frac{n}{J}\right)^2
\exp\left[-\frac{24\pi^2}{cn^2} +\frac{9\sqrt{3}\pi^2}{2c n^{5/2}}
+\left(\frac{\Lambda_n}{\Lambda_c}\right)^{2/3}\right]
\ee
where we
have ignored $\left(\Lambda_{n+1}/\Lambda_c\right)^{2/3}$ as much less
than $\left(\Lambda_n/\Lambda_c\right)^{2/3}$.  The ratio of the
middle term in the exponent to the first term is
$3\sqrt{3}/(16\sqrt{n})\approx 0.05$ for $n\sim 33$, so we will ignore
the middle term.

For sufficiently small $n$, the last term in the exponent dominates
and $P_\obs(n+1)/P_\obs(n)\gg 1$.  There is only an infinitesimal
probability that we will be in a vacuum of level $n$, because there
are others that are much more probable.  As we increase $n$,
$P_\obs(n)$ will continue to increase.  What happens next depends on
the magnitude of $c$.

\subsection{Small $c$}

First suppose $c$ is small, in particular that
\be
\label{eqn:csmall}
c < \frac{24\pi^2}{n^{2/3}J^{4/3}}
\ee
for relevant values of $n$.
For $J = 800$, $n\sim 33$, the right hand side is about $3\times
10^{-3}$.  From Eq.\ (\ref{eqn:csmall}),
\be
\label{eqn:csmall2}
\frac{24\pi^2}{cn^{2/3}J^{4/3}}> 1\,.
\ee
Now
$\Lambda_{n+1}/\Lambda_n\approx (n/J)^2$, and so successive values
of $\Lambda_n^{2/3}$ differ by a factor about $n^{4/3}/J^{4/3}$.

Thus we can find a value of $n$ such that \be \label{eqn:ll} 1 <
\frac{24\pi^2}{cn^{2/3}J^{4/3}}<
\left(\frac{\Lambda_n}{\Lambda_c}\right)^{2/3} <\frac{24\pi^2}{cn^2}
\ee We will now show that for this $n$, \be \label{eqn:PvsP}
P_\obs(n+1)\ll P_\obs(n)\,, \ee so that we should find ourselves in
a vacuum of at most level $n$.

It is not clear from Eq.\ (\ref{eqn:ll}) whether
$\Lambda_{n+1}/\Lambda_c$ is more or less than 1, so we might need
to use either Eq.\ (\ref{eqn:Pobs1}) or Eq.\ (\ref{eqn:Pobs2}) for
$P_\obs(n+1)$.  We will prove the claim using Eq.\
(\ref{eqn:Pobs1}). Since this gives a larger value than Eq.\
(\ref{eqn:Pobs2}), if Eq.\ (\ref{eqn:PvsP}) holds using Eq.\
(\ref{eqn:Pobs1}), it will certainly hold using Eq.\
(\ref{eqn:Pobs2}).  Thus we will take
\be
\label{eqn:pp2}
\frac{P_\obs(n+1)}{P_\obs(n)} \approx
\frac{3\sqrt{\pi}\Lambda_c}{4\Lambda_{n+1}}
\left(\frac{n}{J}\right)^2 \exp\left[-\frac{24\pi^2}{cn^2}
+\left(\frac{\Lambda_n}{\Lambda_c}\right)^{2/3}\right]
\ee

Now from Eq.\ (\ref{eqn:csmall2}), we find that
\be
\frac{24\pi^2}{cn^2}> \left(\frac{J}{n}\right)^{4/3}
\ee
For $J = 800$, $n\sim 33$, the right hand side is about 70.  Thus
unless $(\Lambda_n/\Lambda_c)^{2/3}$ is extremely close to the upper
bound in Eq.\ (\ref{eqn:ll}), the exponential term in Eq.\
(\ref{eqn:pp2}) will be infinitesimal, and Eq.\ (\ref{eqn:PvsP}) will
follow.  If we do have $(\Lambda_n/\Lambda_c)^{2/3}\approx
24\pi^2/(cn^2)$, then $\Lambda_{n+1}/\Lambda_c\agt 1$.  Then the
prefactors in Eq.\ (\ref{eqn:pp2}) are at most about $(n/J)^2$, about
$2\times 10^{-3}$ for parameters of interest, and again Eq.\
(\ref{eqn:PvsP}) follows.

Thus we can say with great confidence that we live in a universe
with level $n$ or lower.  From Eq.\ (\ref{eqn:ll}), we see
immediately that we should observe $\Lambda\ge\Lambda_c$, whereas in
fact we observe $\Lambda_0\approx 0.1\Lambda_c$.  If $c$ is
significantly smaller than the limit in Eq.\ (\ref{eqn:csmall}),
then we will see a very ``non-anthropic'' universe.  We will be able
to find $\Lambda_n$ with $(\Lambda_n/\Lambda_c)^{2/3}>
24\pi^2/(cn^{2/3}J^{4/3})$, and thus \be n_\obs(\Lambda) \alt
n_\obs(\Lambda_n) < e^{-24\pi^2/(cn^{2/3}J^{4/3})} \ee will be tiny,
meaning that only an infinitesimal fraction of matter has coalesced
into galaxies.  For example, with $c=10^{-3}$, we would find
$n_\obs(\Lambda)\alt e^{-3}\approx 0.05$, in contrast to the
observed value (in our approximation) $n_\obs(\Lambda)\approx 0.85$.
With $c=10^{-4}$, we would find $n_\obs(\Lambda)\alt e^{-30}\approx
10^{-13}$: a universe utterly unlike our own.

\subsection{Large $c$}

Now suppose instead that \be c>24\pi^2/n^2 \ee For $J = 800$, $n\sim
33$, the right hand side is about $0.2$.  Then we will reach the
point where $\Lambda_n\sim\Lambda_c$ before $24\pi^2/(cn^2)$ (or the
variation due to $\kappa_{st}$) is significant.  In that case, we
switch to the regime of Eq.\ (\ref{eqn:Pobs1}), where $P_\obs$
decreases only slowly with increasing $n$.  In this regime, \be
\frac{P_\obs(n+1)}{P_\obs(n)}\approx \frac{(J-n)^2}{(J+n+1)^2}
\approx \left(1-\frac{n}{J}\right)^4\,, \ee which is about $0.85$
for parameters of interest.  Thus we find that several values of $n$
contribute nearly equally to the total probability.  The first of
these might be dominated by a single $\Lambda_n$, but the others
will have a large number of closely spaced $\Lambda$.  These
vacua have similar $n_\obs$ and identical prior probability, so we
could easily be in any of them.\footnote{The identical prior
probabilities are a toy feature of the model.  But even if we were
to distribute these probabilities between $P_n$ and $P_{n+1}$, we
would still find many vacua with similar probabilities. If level $n$
has 1 vacuum that is not significantly suppressed by $n_\obs$, then
level $n+1$ will have $N_{n+1}/N_n$ such vacua.  These vacua will be
distributed in a range of probabilities with $P_n/P_{n+1}\sim
N_{n+1}/N_n$, so it is not possible to have a single one of them
strongly dominant.}

Thus when $c$ is large, we recover approximately the original
anthropic predictions with a smooth prior $P(\Lambda)$.  There might
be an effect due to the discrete nature of the vacua associated with
the smallest $n$, where $P_\obs(n)$ has its peak, but this effect is
small because level $n$ does not dominate the probability
distribution.  Instead the probability is divided across many
different levels, while only level $n$ has the above effect.

\section{Discussion}\label{discussion}

A key ingredient in the anthropic prediction of the cosmological
constant is the assumption of a flat prior distribution.  However,
the first attempt to calculate this distribution for the
Bousso-Polchinski and Arkani-Hamed-Dimopolous-Kachru landscape
models \cite{SPV,SP} revealed a staggered distribution, suggesting a
conflict with anthropic predictions.

These calculations have been constrained by computational
limitations and reveal only the probabilities of a handful of the
most probable vacua\footnote{It is extremely unlikely that any of
these vacua should lie in the anthropic range.}. In this paper we
have gone beyond these first order perturbative results by studying
a simple toy model which permits analytic calculation with a large,
realistic number of vacua, $N \sim 10^{500}$. We have found an
interesting fractal distribution for the prior $P(\Lambda)$. When
including anthropic selection effects to determine
$P_{obs}(\Lambda)$, we find that agreement with observation depends
on the only free parameter of the model, the jump size $c$.

We have shown that when $c \sim 1$, anthropic reasoning does indeed
solve the cosmological constant problem.  Even though the prior
distribution has a rich fractal structure, the states of interest
have similar vacua sufficiently closely spaced to approximate the
flat distribution well enough to give the usual anthropic results.

Bousso and Yang \cite{BY} discuss the probability distribution
resulting from the pocket-based measure of Garriga \emph{et al}
\cite{GSPVW}. They claim that it cannot solve the cosmological
constant problem, because the Shannon entropy, $S = -\sum p\ln p$,
computed from the prior probabilities will never obey $\exp(S)\gg
10^{120}$. We feel that a better test is to compute the entropy
using the same formula with the probabilities taking into account
$n_\obs$, and then to demand that $S \gg 1$, so the effective number
of places in the landscape in which we might find ourselves is
large.  This condition is clearly obeyed in the case where $c \sim
1$.

On the other hand, if $c$ is small, of order $10^{-3}$ for $J = 800$,
then the agreement with observation breaks down.  In this case, we
should expect to find ourselves in a universe with quite a large
cosmological constant.  Even though the anthropic factor strongly
disfavors such universes, their volume fraction is so much higher that
in the overall probability they are greatly preferred.

Of course these results apply directly to our toy model with nearly
identical jumps, and there is no reason that the real landscape of
string theory should have this property.  What would happen in a
more realistic theory, which would have many different-sized jumps
in $\Lambda$?

Consider first what happens when we rescale our theory by changing all
jump sizes by a constant factor.  Suppose, for example, that all
different jump sizes are proportional to a parameter $c$.\footnote{For
example,consider a model with $|\Delta\Lambda_i |= c i $ where $1 \leq
i \leq 2J$. $2J$ is the number of directions and $c$ is still some
overall scale.}  The transition rates between vacua all have terms
proportional to $1/c$ in the exponent.  Thus when $c$ is small, the
transition rates and thus the probabilities of different vacua are
very sensitive to the details of the transition rates.  When $c$ is
large, all rates are larger and less variable.  Thus we conjecture
that, as a general rule, landscapes with large jumps are more likely
to give the standard anthropic results, while those with small jumps
are likely to predict universes unlike ours.

Regardless of the details of the theory, there is always a factor
proportional to $\exp(24\pi^2/\Lambda_{\text{max}})$, where
$\Lambda_{\text{max}}$ is the maximum cosmological constant reached in
the series of vacua between the dominant vacuum and ours. If it is
necessary to jump up to a high value of $\Lambda_{\text{max}}$ in
order to have a large enough number of possibilities to expect any
vacuum near the anthropic range, then the exact value of
$\Lambda_{\text{max}}$ will not be so important.  If, however, enough
vacua can be reached with very small $\Lambda_{\text{max}}$, then this
factor will depend strongly on their exact values of
$\Lambda_{\text{max}}$.

If this effect is the dominant one (which is not clear in a realistic
model), then what matters is not the average size of the jumps, but
the number of small jumps.  If there are enough small jumps to produce
some near-anthropic vacua, then these vacua will be preferred, because
of their small $\Lambda_{\text{max}}$, and there may be disagreement
with observation.  The existence of large jumps in addition is then
not important.  Work is underway to study models with different
jump sizes.

\appendix
\section{Bubble abundances by perturbation theory}
As shown in Ref.~\cite{GSPVW}, the calculation of bubble abundances
$p_j$ reduces to finding the smallest eigenvalue $q$ and the
corresponding eigenvector $s$ for a huge $N\times N$
recycling transition matrix $\mathbf{R}$.  Bubble abundances are
given by
\be
p_j\propto \sum_\alpha H_\alpha^q \kappa_{j\alpha} s_\alpha
\approx \sum_\alpha \kappa_{j\alpha} s_\alpha. \label{pJaume}
\ee
where the summation is over all recyclable vacua which can directly
tunnel to $j$.  $H_\alpha$ is the Hubble expansion rate in vacuum
$\alpha$ and we take $H_\alpha^q \approx 1$ because $q$ is an
exponentially small number.

In a realistic model, we expect $N$ to be very large. In the numerical
example of Ref.~\cite{SPV} $N\sim 10^7$, while for a realistic string
theory landscape we expect $N \sim 10^{500}$
~\cite{Susskind,Douglas,AshokDouglas,DenefDouglas}. Solving for the
dominant eigenvector for such huge matrices is numerically impossible.
However, in Ref.~\cite{SPV,SPthesis} the eigenvalue problem was solved via
perturbation theory, with the upward transition rates (see
Eq.~(\ref{eqn:updown})) playing the role of small expansion
parameters. Here we extend this procedure to all orders of
perturbation theory.

We represent our transition matrix as a sum of an unperturbed matrix
and a small correction,
\be
\mathbf{R}=\mathbf{R^{(0)}}+
\mathbf{R^{(1)}},
\ee
where $\mathbf{R^{(0)}}$ contains all the
downward transition rates and $\mathbf{R^{(1)}}$ contains all the
upward transition rates.  We will solve for the zero'th order
dominant eigensystem $\{q^{(0)},\mathbf{s^{(0)}}\}$ from
$\mathbf{R^{(0)}}$ and then include contributions from
$\mathbf{R^{(1)}}$ to all orders of perturbation theory.

If the vacua are arranged in the order of increasing $\Lambda$, so
that
\be
\Lambda_1 \leq \Lambda_2 \leq \ldots \leq \Lambda_N,
\ee
then $\mathbf{R^{(0)}}$ is an upper triangular matrix.  Its
eigenvalues are simply equal to its diagonal elements,
\be
R^{(0)}_{\alpha\alpha} =-\sum_{j<\alpha}\kappa_{j\alpha} \equiv
-D_\alpha. \label{Ralpha}
\ee
Hence, the magnitude of the smallest
zeroth-order eigenvalue is
\be
q^{(0)}=D_{{*}} \equiv {\rm
min}\{D_\alpha\}. \label{q0}
\ee

Downward transitions from ${*}$ will bring us to the
negative-$\Lambda$ territory of terminal vacua \cite{SPV}.  Terminal
vacua do not belong in the matrix $\mathbf{R}$; hence,
$R_{\beta{*}}=0$ for $\beta\neq{*}$, and we see that the zeroth order
eigenvector has a single nonzero component,
\be
s^{(0)}_\alpha=
\delta_{\alpha{*}}. \label{s0}
\ee
Thus, in fact, we could have
included only the diagonal elements of $\mathbf{R^{(0)}}$ and still
obtained the correct $q^{(0)}$ and $\mathbf{s^{(0)}}$.

Now we would like to include the effect of the upward transition rates
in the lower triangular matrix $\mathbf{R^{(1)}}$, to any given order
of perturbation theory.  To organize the calculation, we note that
the eigenvalues of an upper triangular matrix are just the diagonal
elements, and the eigenvectors are given exactly by the perturbation
series, which terminates after $N$ terms.  Thus we will take
just the diagonal elements of $\mathbf{R^{(0)}}$ as our unperturbed
matrix, and consider the rest of $\mathbf{R^{(0)}}$ and
$\mathbf{R^{(1)}}$ as a perturbation.  By summing all terms in the
perturbation series for this perturbation that involve $n$ elements of
$\mathbf{R^{(1)}}$, we find the perturbation term of order $n$ in the
small upward jump rates.

The procedure is the same as that involved in finding an eigenstate of
the Schr\"odinger equation in $n$th-order perturbation theory, working
in a basis of wavefunctions which diagonalize the unperturbed
equation.  See for example Eq.~(9.1.16) of Ref.~\cite{MorseFeshbach}.
Including all orders of perturbation theory, we find the result
\be
\label{eqn:pfinal}
s_a = \sum_{b=1}^N\cdots\sum_{z=1}^N
\frac{\kappa_{ab}}{(D_a-D_*)}
\cdots \frac{\kappa_{z*}}{(D_z-D_*)} + \text{additional terms}
\ee
where there can be any number of terms in the sum and the vacuum * is
not summed over.

The ``additional terms'' are those which have products of at least two
elements of the perturbation matrix with index *.  We can write such a
term as a sequence of transitions which returns to the vacuum *
(perhaps several times) before finally going on to the vacuum $a$.  By
keeping only the final part of the path from * to $a$, we get a
related term which occurs in the sum in Eq.\ (\ref{eqn:pfinal}).  The
deleted part of the path must contain at least one matrix element of
the form $\kappa_{i*}$, which is an upward jump, or if both indices
are *, the matrix element (called $U_*$ in~\cite{SPV}) is the sum of
all upward jump rates from *.  Since all upward jumps are highly
suppressed, the additional term is tiny compared to the related term
in the sum in Eq.\ (\ref{eqn:pfinal}).  So, ignoring the additional
terms, Eq.\ (\ref{eqn:pfinal}) gives each $s_a$ correct at first
nonvanishing order in the number of upward jumps.

Combining Eqs.\ (\ref{eqn:pfinal}) and (\ref{pJaume}) gives
Eq.\ (\ref{eqn:pl}).

\section*{Acknowledgments}
We would like to thank Alex Vilenkin for many key discussions and
ideas, and Claire Zukowski for pointing out the need for the
additional terms in Eq.\ (\ref{eqn:pfinal}).  K.D.O. was supported in
part by the National Science Foundation under grant 0353314.
D. S.-P. was supported in part by grant RFP1-06-028 from The
Foundational Questions Institute (fqxi.org).

\end{document}